\documentclass[preprint,amsmath]{revtex4}
\usepackage{CJK}
\usepackage{amsmath}
\usepackage{enumerate}
\usepackage{graphicx}
\usepackage{amsfonts}
\usepackage{url}
\usepackage{bm}
\usepackage{tikz}  
\usepackage{pgf}   

\usepackage{pifont}
\usepackage{epsfig,subfigure,dsfont,amsthm,amsbsy,mathrsfs,amscd}
\newtheorem{theorem}{Theorem}

\newtheorem{definition}{Definition}

\newtheorem{proposition}{Proposition}

\begin{document}

\begin{center}
\bf {The uncertainty of quantum states with respect to the projective measurement}
\end{center}

\begin{center}
{Ming-Jing Zhao$^{1,\ast}$,  Yuanhong Tao$^{2,\dagger}$
\vspace{2ex}

{\small $^1$School of Science, Beijing Information Science and Technology University, Beijing, 100192, P. R. China\\}
{\small $^2$Department of Big Data, School of Science, Zhejiang University of Science and Technology, Hangzhou 310023, P. R. China}}

\end{center}

{{\bf Abstract}
The uncertainty relation is a distinctive characteristic of quantum theory. The uncertainty is essentially rooted in quantum states. In this work we regard the uncertainty as an intrinsic property of quantum state and characterize it systematically with respect to given projective measurement. Some basic concepts about uncertainty are reformulated in this context. We prove and get the form of the uncertainty preserving operations. The quantum states with maximal uncertainty are characterized.
A universal decomposition of uncertainty into classical uncertainty and quantum uncertainty is provided.
Furthermore, a unified and general relation among uncertainty, coherence and coherence of assistance is established. These results are independent of any explicit uncertainty measure.
At last, we propose a new uncertainty measure called the geometric uncertainty based on the fidelity and link it with the geometric coherence.
}

{\bf Emails:}{$^\ast$} zhaomingjingde@126.com,
$^\dagger$ taoyuanhong12@126.com


\section{Introduction}

The uncertainty relation is first introduced by Heisenberg
to describe the restrictions on the accuracy of measurement results of two
or more noncommutative observables \cite{W. Heisenberg,H. P. Robertson}.
Formally the uncertainty relations are established in forms of variance \cite{H. P. Robertson}, Shannon entropy \cite{D. Deutsch,H. Maassen,Coles. P.J.},
R\'{e}nyi entropy \cite{H. Maassen},
conditional entropy \cite{Renes. J.M,Gour. G}, mutual information \cite{A. Grudka}, etc.
A universal uncertainty relation is then discovered with majorization techniques \cite{S. Friedland,Pucha}.
As the applications, the uncertainty relation is factually the central ingredient in
the security analysis of almost all quantum cryptographic protocols, such as quantum key distribution
and two-party quantum cryptography \cite{Tomamichel. M.}.
The uncertainty relation has also been shown to be a powerful tool in
the quantum random number generation \cite{G. Vallone,Z. Cao}, entanglement witness \cite{M. Berta}, EPR steering
\cite{S. P. Walborn,J. Schneeloch}, and quantum metrology \cite{V. Giovannetti}.

However one fundamental question is where does the uncertainty stem from? It is from quantum states. In other words, the uncertainty is an intrinsic property of quantum states. Since it is displayed by means of the measurement results of observables, so it also relied on the observables.
For example, under the projective measurement $\Pi=\{\Pi_i\}_{i=0}^{d-1}$ with $\Pi_i=|i\rangle\langle i|$, the uncertainty of quantum state $\rho$ is reflected by the probability vector $(\langle 0|\rho|0\rangle, \langle 1|\rho|1\rangle,\cdots, \langle d-1|\rho|d-1\rangle)^T$. More specifically, let us consider the mixed state $\rho=\frac{1}{2}(|0\rangle\langle0|+|1\rangle\langle1|)$ and pure state $|\psi\rangle=\frac{1}{\sqrt{2}}(|0\rangle + |1\rangle)$ in qubit system. Under the projective measurement $\{\Pi_i\}=\{|i\rangle\langle i|\}_{i=0,1}$, the measurement results are ${\rm Tr}(\Pi_i \rho \Pi_i)=\langle \psi|\Pi_i| \psi\rangle=\frac{1}{2}$ for $i=0,1$. Therefore the uncertainties existed in $\rho$ and $|\psi\rangle$ are equal under the the projective measurement $\{\Pi_i\}$.

In fact, the uncertainty of quantum states is manifested in two forms, classical uncertainty and quantum uncertainty \cite{S. Luo2005,S. Luo2005-1,S. Luo2017,Y. Sun2021-1,Y. Sun2021}. The classical uncertainty arises from classical mixing while the quantum uncertainty arises from the superposition. In this sense, the
uncertainty in mixed state $\rho=\frac{1}{2}(|0\rangle\langle0|+|1\rangle\langle1|)$ is classical uncertainty and the uncertainty in pure state $|\psi\rangle=\frac{1}{\sqrt{2}}(|0\rangle + |1\rangle)$ is quantum uncertainty \cite{S. Luo2005,S. Luo2017,Y. Sun2021-1,Y. Sun2021}. Generally speaking, the uncertainty is an abundant resource existed in quantum state.

However, compared with other quantumness like entanglement \cite{Horodecki}, correlation \cite{Modi} and coherence \cite{revStreltsov,revHu,Marvian,Hillery}, the study of uncertainty in quantum states is rather inadequate. Our aim is to characterize the uncertainty in quantum states unambiguously and formally and link the uncertainty with other quantumness tightly. Here we fix the projective measurement $\Pi=\{\Pi_i\}_{i=0}^{d-1}$ with $\Pi_i=|i\rangle\langle i|$ and explore the uncertainty of quantum states under this measurement systematically.

The remainder of the paper is arranged as follows. In section II, some basic concepts including the free states, free operations and measures about uncertainty in the framework of resource theory are presented. Here we reformulate the uncertainty in terms of the intrinsic property of quantum state with respect to the projective measurement. We show a necessary and sufficient condition for the uncertainty preserving operation. In section III, we characterize the quantum states with maximal uncertainty. In section IV, we show a universal decomposition of uncertainty into classical uncertainty and quantum uncertainty both formally and quantitatively. In section V, we establish a unified and general relation among uncertainty, coherence and coherence of assistance quantitatively. In section VI, we recall two uncertainty measures and propose a new uncertainty measure named geometric uncertainty. The decompositions of uncertainty into classical part and quantum part are also made respectively. In section VII, we conclude with some outlooks that remain to be answered in uncertainty theory.

\section{The uncertainty of quantum states}

For any $d$-dimensional Hilbert space $\mathcal{H}_{d}$, suppose $\{|i\rangle\}_{i=0}^{d-1}$ is any orthonormal basis of $\mathcal{H}_{d}$. For any quantum state $\rho$ in $\mathcal{H}_{d}$,
we refer to the uncertainty of the quantum state $\rho$ as the uncertainty of the probability vector
$(\langle 0|\rho|0\rangle, \langle 1|\rho|1\rangle,\cdots, \langle d-1|\rho|d-1\rangle)^T$ with respect to the projective measurement
$\Pi=\{\Pi_i\}_{i=0}^{d-1}$ with $\Pi_i=|i\rangle\langle i|$ throughout this paper. In this sense the uncertainty of quantum state is basis-dependent.
In the framework of the uncertainty resource theory, we specify the free states as certain states and free operations as certain operations. Next we characterize the free states and free operations and present a necessary and sufficient condition for free operations.

First, for any quantum state $\rho$, we call it {\it certain state} as the free state in the uncertainty resource theory if $\rho$ is rank-one diagonal under the reference basis, i.e. $\rho=|i\rangle\langle i|$ for $i=0,1,\cdots, d-1$. Or else it is uncertain.
Let $\mathcal{C}$ denote the set of certain states, $\mathcal{C}=\{|0\rangle\langle 0|,|1\rangle\langle 1|, \cdots, |d-1\rangle\langle d-1|\}$. All quantum states are classified into certain states and uncertain states. The uncertain states contains uncertainty, that is classical uncertainty or quantum uncertainty or both \cite{S. Luo2005,S. Luo2017,Y. Sun2021-1,Y. Sun2021}.
For pure states $|\psi\rangle=\sum_i \psi_i |i\rangle$, it has no classical uncertainty and its uncertainty is attributed to the quantum uncertainty. For incoherent states which is diagonal under the reference basis, $\rho=\sum_i \rho_{ii} |i\rangle\langle i|$, it has no quantum uncertainty and its uncertainty is attributed to the classical uncertainty. For general coherent mixed states, it contains both the classical uncertainty and quantum uncertainty.

Furthermore, for any quantum operation $\Lambda$, we call it {\it certain operation} as the free operation in the uncertainty resource theory  if $\Lambda(\rho)$ keeps certain for all certain quantum states $\rho$. We show the certain operations are in the following form.

\begin{theorem}\label{th certain operation}
All certain operations $\Lambda$ are in form with
Kraus representation $\Lambda=\sum_l K_l (\cdot) K_l^\dagger$ with $K_l=\sum_i \sqrt{p_{il}} e^{{\rm i}\theta_{il}} |g(i)\rangle \langle i|$, $\sum_l p_{il}=1$ for all $i$, and $\sum_l  \sqrt{p_{i^\prime l} p_{il}}  e^{{\rm i}(\theta_{il} -\theta_{i^\prime l})}  \langle g(i^\prime)|  g(i)\rangle =0$ for $i\neq i^\prime$, $g$ is a map from $\{i\}_{i=0}^{d-1}$ to $\{i\}_{i=0}^{d-1}$.
\end{theorem}

The proof is in Appendix \ref{app cer}. Theorem \ref{th certain operation} shows that for certain operations, the positions of the nonzero entries of all Kraus operators are the same and there are at most one nonzero entries in each row for every Kraus operator.
Explicitly, all permutation operators $P_{\pi}=\sum_i |\pi(i)\rangle \langle i|$ with $\pi$ any permutation in the permutation group are certain operations. But certain operations may be not permutations. For example, in qubit system, let $K_1=|0\rangle\langle0|$ and $K_2=|0\rangle\langle1|$ satisfying $K_1^\dagger K_1 + K_2^\dagger K_2=I$, then the quantum operation $\Lambda= K_1 (\cdot) K_1^\dagger+K_2 (\cdot) K_2^\dagger$ is also certain because $\Lambda(|0\rangle\langle0|)=\Lambda(|1\rangle\langle1|) =|0\rangle\langle0|$.
For diagonal matrices $D_l$ with $\sum_l D_l^\dagger D_l=I$,
the quantum operations $\Lambda=\sum_l D_l (\cdot)D_l^\dagger$ are also certain operations. Furthermore the composition of the certain operations is still certain.
Therefore for any permutation $P_{\pi}$, the quantum operation $\Lambda=\sum_l P_{\pi} D_l(\cdot)D_l^\dagger P_{\pi}$ is also certain.

For the quantification of the uncertainty, an axiomatic approach has been proposed which requires that
any reasonable measure of uncertainty is a function of the
probability vector satisfying (1) invariant under permutations of its elements, (2) nondecreasing under a random
relabeling of its argument \cite{S. Friedland}.
Mathematically, the uncertainty measure should be a non-negative
Schur-concave function that takes the value zero on
the vector $\vec{x}=(1,0,\cdots,0)^T$ \cite{S. Friedland}. Since all symmetric concave functions are Schur-concave function,
here we restrict to the nonnegative symmetric concave functions [See Appendix \ref{app sym con f}] as the uncertainty measure.

\begin{proposition}\label{U measure}
For any nonnegative symmetric concave function $f$ and any quantum state $\rho$,
$U(\rho)=f(\langle 0|\rho|0\rangle, \langle 1|\rho|1\rangle,\cdots, \langle d-1|\rho|d-1\rangle)$ is an uncertainty measure of $\rho$.
\end{proposition}

For any given uncertainty measure $U$ and any quantum operations $\Lambda$, we call $\Lambda$ {\it uncertainty preserving} if it preserves uncertainty for every quantum state $\rho$, $U(\Lambda(\rho))=U(\rho)$.
The uncertainty preserving operations can be characterized as follows.

\begin{theorem}\label{th uncertainty pres}
For any quantum operation $\Lambda=\sum_l K_l(\cdot)K_l^\dagger$, it is an uncertainty preserving operation if and only if the Kraus operators $K_l= D_lP_{\pi}$ with diagonal matrices $D_l$ such that $\sum_l D_l^\dagger D_l=I$ and some permutation $P_{\pi}$.
\end{theorem}

The proof is in the Appendix \ref{app uncer pre}. The uncertainty preserving operation is factually a special certain operation.
Under the action of uncertainty preserving operations, the uncertainty is kept, but the quantum part does not increase. So the uncertainty preserving operation transfers the quantum uncertainty into classical uncertainty.
For example, the projective measurement $\Pi=\{\Pi_i\}_{i=0}^{d-1}$ with $\Pi_i=|i\rangle\langle i|$ is an uncertainty preserving operation, it projects any quantum state $\rho=\sum_{ij} \rho_{ij} |i\rangle\langle j|$ into diagonal states $\Pi(\rho)=\sum_i \Pi_i \rho \Pi_i =\sum_{i} \rho_{ii} |i\rangle\langle i|$. Initially $\rho$ contains quantum uncertainty and classical uncertainty, after the projective measurement $\Pi$, the resulted state $\Pi(\rho)$ just contains classical uncertainty.

\section{Maximally uncertain states}

In this section, we shall study and characterize the quantum states with maximal uncertainty under the projective measurement $\Pi=\{\Pi_i\}_{i=0}^{d-1}$ with $\Pi_i=|i\rangle\langle i|$.

\begin{definition}
For any quantum state $\rho$ and any uncertainty measure $U$,
$\rho$ is called maximally uncertain if the uncertainty of $\rho$ reaches the maximum of $U(\rho)$ over all quantum states.
\end{definition}

Under the projective measurement $\Pi=\{\Pi_i\}_{i=0}^{d-1}$ with $\Pi_i=|i\rangle\langle i|$, the maximally uncertain states can be characterized definitely.
\begin{theorem}\label{th maxi uncer}
For $d$-dimensional quantum state $\rho=\sum_{ij} \rho_{ij} |i\rangle \langle j|$, it is maximally uncertain if and only if $\rho_{ii}=1/d$ for $i=0,1,\cdots, d-1$.
\end{theorem}

\begin{figure}[ht]\centering
{\begin{minipage}[b]{0.5\linewidth}
\includegraphics[width=1\textwidth]{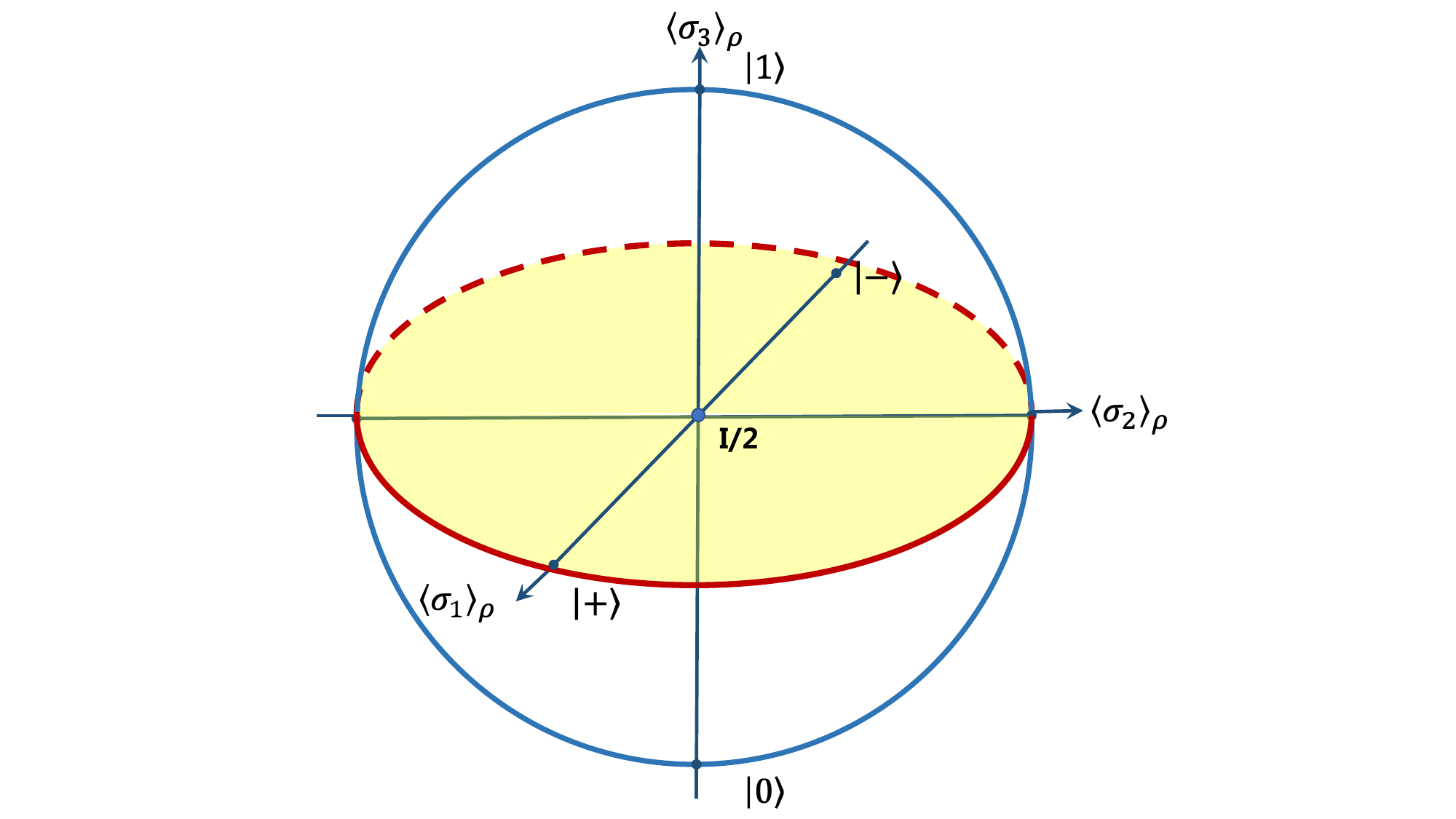}
\end{minipage}}
\caption{(Color Online) The distribution of maximally uncertain states in Bloch sphere.
All maximally uncertain states are distributed in the largest horizontal
disc. The boundary is the quantum states $|\psi\rangle=\frac{1}{\sqrt{2}}(|0\rangle + e^{{\rm i}\theta} |1\rangle)$ with maximal quantum uncertainty and the spherical center is the quantum state $I/2$ with maximal classical uncertainty.}
\label{fig}
\end{figure}

The proof is in Appendix \ref{app maxi uncer}. Theorem \ref{th maxi uncer} shows that although there are many uncertainty measures and the maximums of the uncertainty measures are probably different, the maximally uncertain states are definite.
In a qubit system, all quantum states can be represented in the Bloch sphere. All maximally uncertain states are distributed in the largest horizontal
disc, and the pure states $|\psi\rangle=\frac{1}{\sqrt{2}}(|0\rangle + e^{{\rm i}\theta} |1\rangle)$ with maximal quantum uncertainty are distributed in the largest horizontal circle. The spherical center is the quantum state $I/2$ with maximal classical uncertainty (See Fig. \ref{fig}). Overall the maximally uncertain states are the furthest from the certain states $|0\rangle$ and $|1\rangle$ vertically.

In fact all quantum states can be transformed into maximally coherent states.
For any mixed state $\rho$, let the unitary operator $Q=\sum_{i=0}^{d-1} |i\rangle \langle i+1\mod{d}|$, first, we perform the quantum operation $\Lambda_1=\frac{1}{d} \sum_{t=1}^d Q^{t} (\cdot) Q^{t}$ on quantum state $\rho$ with $Q^{t}$ the power $t$ of $Q$. By directly calculating we get the diagonal entries of $\Lambda_1(\rho)$ are all $1/d$, which means $\Lambda_1(\rho)$ is maximally uncertain. Compared with the initial quantum state $\rho$, the uncertainty has been increased, but the quantum uncertainty is decreased as the operation $\Lambda_1$ is incoherent operation. So the increased uncertainty of $\rho$ is classical uncertainty. Second, we perform the projective measurement $\Pi=\{\Pi_i\}_{i=0}^{d-1}$ with $\Pi_i=|i\rangle\langle i|$ on $\Lambda_1(\rho)$ and get the output state $I/d$ which can be decomposed as the equal weight of the maximally coherent pure states. Third, with the one-way assisted coherence distillation $\Lambda_3$ \cite{E. Chitambar}, the quantum state $I/d$ is transformed into the maximally coherent pure state $|\Psi_d\rangle=\frac{1}{\sqrt{d}} \sum_i |i\rangle$. This process is
\begin{eqnarray*}\label{eq rho to max}
\rho  \xrightarrow{\Lambda_1} \Lambda_1(\rho)  \xrightarrow{\Pi} I/d \xrightarrow{\Lambda_3} |\Psi_d\rangle.
\end{eqnarray*}
Therefore, with these three steps, any quantum state can be transformed into maximally coherent states.

From another point of view, any quantum state $\rho$ is maximally uncertain under some projective measurement $\{\Pi_\alpha\}=\{|\alpha\rangle \langle\alpha|\}$
with $\{|\alpha\rangle\}$ some orthonormal basis of $\mathcal{H}_{d}$.
This is because any matrix is unitarily equivalent to a matrix with constant main
diagonal \cite{fillmore}. Therefore, there exist some unitary matrix $U$ such that the diagonal entries of $U^\dagger \rho U$ are all $1/d$. Then under the
projective measurement  $\{\Pi_\alpha\}=\{|\alpha\rangle \langle\alpha|\}$ consisted by the eigenvectors of $U$, we have $\langle \alpha|\rho|\alpha\rangle=1/d$ for all $\alpha$. So $\rho$ is maximal uncertain under this projective measurement $\{\Pi_\alpha\}$.

\section{A universal decomposition of uncertainty into quantum uncertainty and classical uncertainty}

Generally the uncertainty can be decomposed into quantum uncertainty and classical uncertainty formally,
\begin{equation}\label{eq dec}
\text{Uncertainty=Quantum Uncertainty + Classical Uncertainty.}
\end{equation}
Hence the uncertainty is termed total uncertainty when necessary to avoid any confusion with quantum uncertainty and classical uncertainty.
Originally, quantum uncertainty arises from the superposition while classical uncertainty arises from the classical mixing. So the quantum uncertainty is also recognized as coherence and the classical uncertainty is recognized as mixedness.

For quantum uncertainty, let us recall the coherence theory briefly. Under the fixed reference basis $\{|i\rangle\}$, the quantum state is incoherent if it is diagonal, $\rho=\sum_{i =0}^{d-1} \rho_{ii} |i\rangle\langle i|$. This set of incoherent
states is labeled by ${\cal I}$. The quantum state is called coherent if it is not incoherent. Quantum operation $\Lambda=\sum_l K_l (\cdot) K_l^\dagger $ is an incoherent operation if it fulfills
$K_l\sigma K_l^{\dag}/{\rm Tr}(K_l\sigma K_l^{\dag})\in {\cal I}$ for all
$\sigma\in {\cal I}$ and for all $l$.
Any coherence measure $C(\rho)$ is a convex extension of $C(|\psi\rangle)$ from pure states to mixed states with $C(|\psi\rangle)=f(|\psi_0|^2,|\psi_1|^2,\cdots,|\psi_{d-1}|^2)$,  where $f$ is some nonnegative symmetric concave function, and $|\psi\rangle=\sum_i \psi_i|i\rangle$ \cite{H. Zhu,S. Du}.
Some common used coherence measures are $l_1$ norm coherence \cite{Baumgratz}, relative entropy coherence \cite{Baumgratz}, coherence concurrence \cite{TGao} and so on.
The gap between total uncertainty and quantum uncertainty is classical uncertainty. One well-known classical uncertainty quantifier is linear entropy $S_2(\rho)=1- {\rm Tr}\rho^2$. It is worthy to point out that the uncertainty and quantum uncertainty are both basis dependent, while the classical uncertainty is independent of the measurement basis.

In fact, the total uncertainty can be decomposed quantitatively into quantum uncertainty and classical uncertainty.
For any given nonnegative symmetric concave function $f$ and any quantum state $\rho=\sum_{i,j} \rho_{ij} |i\rangle\langle j|$, we have
\begin{eqnarray}\label{uni de}
U(\rho)=C(\rho)+D(\rho),
\end{eqnarray}
where $U(\rho)=f(\langle 0|\rho|0\rangle, \langle 1|\rho|1\rangle,\cdots, \langle d-1|\rho|d-1\rangle)$ quantifies the total uncertainty existed in $\rho$,
$C(\rho)$ is any convex extension of $C(|\psi\rangle)$ from pure states to mixed states with $C(|\psi\rangle)=f(|\psi_0|^2,|\psi_1|^2,\cdots,|\psi_{d-1}|^2)$ for $|\psi\rangle=\sum_i \psi_i|i\rangle$ and quantifies the quantum uncertainty existed in $\rho$, and $D(\rho)=U(\rho)-C(\rho)$ quantifies the classical uncertainty existed in $\rho$.
Therefore, the equality (\ref{uni de}) gives rise to an operational decomposition
corresponding to the equality (\ref{eq dec}). We should emphasize that
the decomposition (\ref{uni de}) is universal because it applies for all uncertainty measures.
Additionally, the  decomposition (\ref{uni de}) is not unique because the coherence measure $C(\rho)$ as the convex extensions of $C(|\psi\rangle)$ is not unique. The comparison among the total uncertainty, quantum uncertainty and classical uncertainty are made in Table I.

\begin{table}[!h]\label{table}
\newcommand{\tabincell}[2]{\begin{tabular}{@{}#1@{}}#2\end{tabular}}
\begin{center}
\def\temptablewidth{1\textwidth}
{\rule{\temptablewidth}{2pt}}
\begin{tabular*}
{\temptablewidth}{@{\extracolsep{\fill}}c|c|c|c}
& Total uncertainty $U(\rho)$  & Quantum uncertainty $C(\rho)$  &  Classical uncertainty $D(\rho)$  \\   \hline
Nullity & Certain states $|i\rangle\langle i|$ & Incoherent states $\sum_i \rho_{ii}|i\rangle\langle i|$ & Pure states  \\   \hline
Maximizer &  $\rho=\sum_{i,j} \rho_{ij} |i\rangle\langle j|$ with $\rho_{ii}=1/d$ &  $\frac{1}{\sqrt{d}}\sum_i e^{{\rm i}\theta_i} |i\rangle$ & $I/d$  \\  \hline
Convexity & Concave  &  Convex & Concave
\end{tabular*}
{\rule{\temptablewidth}{2pt}}
 \caption{The comparison among the total uncertainty, quantum uncertainty and classical uncertainty in $d$-dimensional systems.
  }
 \end{center}
 \end{table}

\section{Uncertainty, coherence and coherence of assistance}

For any coherence measure $C$, there is another quantifier called
the coherence of assistance $C_a$ corresponding to it, which is defined as
\begin{eqnarray}\label{ca}
C_a(\rho)=\max_{\{p_k,|\psi_k\rangle\}} \sum_k p_k C(|\psi_k\rangle)
\end{eqnarray}
with the maximum running over all possible pure state decompositions of $\rho=\sum_k p_k |\psi_k\rangle \langle\psi_k|$. This is the concave bottom extension of $C(|\psi\rangle)$ from pure states to mixed states. It can be interpreted
operationally in the following way.
Suppose Alice holds a state
$\rho^A$ with coherence $C(\rho^A)$. Bob holds another part of the
purified state of $\rho^A$. The joint state between Alice and Bob is
$\sum_k p_k |{\psi_k}\rangle_A\otimes |{k}\rangle_B$ with $\rho^A=\sum_k
p_k |{\psi_k}\rangle\langle \psi_k|$.
Bob performs local measurements $\{|{k}\rangle\}$
and informs Alice the measurement outcomes by classical
communication. Alice's quantum state will be in a pure state
ensemble $\{ p_k, |{\psi_k}\rangle\}$ with average coherence $\sum_k
p_k C(|{\psi_k}\rangle\langle \psi_k|)$. This process is a special case of assisted coherence distillation to generate the maximal possible coherence on the target system in bipartite systems by only incoherent operations on the target system and local quantum operations on the auxiliary system. It enables Alice to increase the
coherence from $C(\rho^A)$ to the average coherence $\sum_k p_k
C(|{\psi_k}\rangle\langle \psi_k|)$ due to the convexity of the coherence measure. The maximum average coherence is known as the coherence of assistance \cite{E. Chitambar}.

For any given nonnegative symmetric concave function $f$ and any quantum state $\rho=\sum_{i,j} \rho_{ij} |i\rangle \langle j|$, any convex extension of $f(|\psi_0|^2,|\psi_1|^2,\cdots,|\psi_{d-1}|^2)$ for any pure state $|\psi\rangle=\sum_i \psi_i |i\rangle$ to mixed states gives rise to a coherence measure $C(\rho)$ and the concave bottom extension in Eq. (\ref{ca}) gives rise to the corresponding coherence of assistance $C_a(\rho)$. Since we can also derive an uncertainty measure by $U(\rho)=f(\rho_{00}, \rho_{11}, \cdots, \rho_{d-1,d-1})$, so
the uncertainty measure $U$, coherence measure $C$ and the coherence of assistance $C_a$ are connected by the real symmetric concave function $f$ (See Fig. 2).
Furthermore, they are also comparable quantitatively.

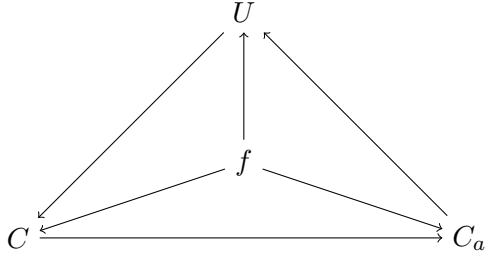
\begin{figure}\label{figf}
\begin{tikzpicture}[->=stealth,auto,node distance=2cm]
\centering
\node (f) 
{$f$};
\node (ca) [xshift=1cm,yshift=-1cm,right of =f] 
{$C_a$};
\node (cc) [xshift=-1cm,yshift=-1cm,left of =f] 
{$C$};
\node (u) [xshift=0cm,yshift=2cm] 
{$U$};

\draw[->] (f) -- (cc) ;  
\draw[->] (f) -- (ca) ;
\draw[->] (f) -- (u) ;
\draw[->] (cc) -- (ca) ;
\draw[->] (ca) -- (u) ;
\draw[->] (u) -- (cc) ;
\end{tikzpicture}
\caption{Here $f$ is any nonnegative symmetric concave function. It deduces uncertainty measure $U$, coherence measure $C$ and coherence of assistance $C_a$. The quantum part of uncertainty $U$ is the coherence $C$. By local operation and one-way classical communication with the assistance of another party, the coherence $C$ can be increased into $C_a$, which is no more than the uncertainty $U$.}
\end{figure}

\begin{theorem}\label{th relation}
For any quantum state $\rho$, the relation among its uncertainty, coherence, and coherence of assistance is
\begin{eqnarray}
C(\rho) \leq C_a(\rho)\leq U(\rho).
\end{eqnarray}
\end{theorem}

The proof is in Appendix \ref{app rel}.
In the process of assisted coherence distillation above, the quantum coherence can be increased.
It lies in that the classical uncertainty of mixed state $\rho$ is transformed not necessary totally into the quantum uncertainty with the help of Bob in the process of purification. Since the purification does not affect the total uncertainty, the coherence of assistance can not beyond the total uncertainty.
In fact the coherence of assistance in qubit and qutrit systems can reach the total uncertainty, that is, $C_a(\rho)=U(\rho)$ for two and three dimensional quantum states \cite{M. Zhao}. But it fails in high dimensional systems.

\section{Three uncertainty measures}

In this section, we consider three uncertainty measures which correspond to the nonnegative symmetric concave functions as (1) $f(\rho_{00}, \rho_{11},\cdots, \rho_{d-1,d-1})=1-\sum_i \rho_{ii}^2$; (2) $f(\rho_{00}, \rho_{11},\cdots, \rho_{d-1,d-1})=-\sum_i \rho_{ii}\log\rho_{ii}$; and (3) $f(\rho_{00}, \rho_{11},\cdots, \rho_{d-1,d-1})=1-\max_i \rho_{ii}$. They are originated from the variance, entropy and fidelity respectively.

{\it{Example 1. The uncertainty measure of quantum states based on variance with $f(\rho_{00}, \rho_{11},\cdots, \rho_{d-1,d-1})=1-\sum_i \rho_{ii}^2$.}}

Under the projective measurement $\Pi=\{\Pi_i\}_{i=0}^{d-1}$ with $\Pi_i=|i\rangle \langle i|$,
the total uncertainty in quantum state $\rho=\sum_{i,j} \rho_{ij} |i\rangle\langle j|$ in terms of variance is
\begin{eqnarray}
U_{var}(\rho)=\sum_i V(\Pi_i,\rho)
=1-\sum_i \rho_{ii}^2,
\end{eqnarray}
where the variance of observable $A$ in the quantum state $\rho$ is $V(A,\rho)={\rm Tr}(A^2\rho)-[{\rm Tr}(A\rho)]^2$.
The total uncertainty $U_{var}(\rho)$ can be decomposed as \cite{S. Luo2005,Y. Sun2021}
\begin{eqnarray}
U_{var}(\rho)=C(\rho)+I(\rho),
\end{eqnarray}
where
$I(\rho)=\sum_i I(\rho, \Pi_i)=\sum_i -\frac{1}{2} {\rm Tr} [\sqrt{\rho}, \Pi_i]^2=1-\sum_i \langle i| \sqrt\rho |i\rangle^2  $ is the quantum uncertainty quantified by the skew information and $C(\rho)=[\sum_i (\langle i| \sqrt\rho |i\rangle^2-\rho_{ii}^2)]$ is the classical uncertainty.
The total uncertainty $U_{var}(\rho)$ can be also decomposed as
\begin{eqnarray}
U_{var}(\rho)=S_2(\rho)+Q(\rho),
\end{eqnarray}
where $S_2(\rho)=1- {\rm Tr}\rho^2$ is the classical uncertainty quantified by the linear entropy and $Q(\rho)= S_2(\Pi(\rho))- S_2(\rho)$ with $\Pi(\rho)=\sum_i \Pi_i \rho \Pi_i$ is the quantum uncertainty \cite{Y. Sun2021-1}.

{\it{Example 2. The uncertainty measure of quantum states based on entropy with $f(\rho_{00}, \rho_{11},\cdots, \rho_{d-1,d-1})=-\sum_i \rho_{ii}\log\rho_{ii}$.}}

Under the projective measurement $\Pi=\{\Pi_i\}_{i=0}^{d-1}$ with $\Pi_i=|i\rangle \langle i|$, the total uncertainty in quantum state $\rho=\sum_{i,j} \rho_{ij} |i\rangle\langle j|$ in terms of entropy is
\begin{eqnarray}
U_s(\rho)=S(\Pi(\rho))=-\sum_i \rho_{ii}\log\rho_{ii},
\end{eqnarray}
which can be decomposed as
\begin{eqnarray}
U_s(\rho)=S(\rho) + C_r(\rho)
\end{eqnarray}
with entropy $S(\rho)$ quantifying the classical uncertainty and the relative entropy $C_r(\rho)=S(\rho\|\Pi(\rho))=S(\Pi(\rho))-S(\rho)$ quantifying the quantum uncertainty \cite{T. Baumgratz,X. Yuan2017}.

{\it{Example 3. The uncertainty measure of quantum states based on fidelity with $f(\rho_{00}, \rho_{11},\cdots, \rho_{d-1,d-1})=1-\max_{i} \rho_{ii}$.}}

Now we propose a new uncertainty measure based on the fidelity $F(\rho,\sigma)=[{\rm Tr}(\sqrt{\rho} \sigma \sqrt{\rho})^\frac{1}{2}]^2$ \cite{Jozsa}.
Let
\begin{eqnarray}
U_f(\rho)=1-\max_{\sigma\in\mathcal{C}} F(\rho,\sigma),
\end{eqnarray}
where the maximum is taken over all certain states. We call it geometric uncertainty and show it is a well-defined uncertainty measure.

\begin{theorem}\label{th uf}
For any quantum state $\rho=\sum_{i,j} \rho_{ij} |i\rangle\langle j|$, the geometric uncertainty $U_f(\rho)$ is an uncertainty measure and
\begin{eqnarray}
U_f(\rho)=1-\max_{i} \rho_{ii}.
\end{eqnarray}
\end{theorem}

The proof is in Appendix \ref{app uncer fie}.
The geometric uncertainty $U_f$ has the following properties.
\begin{enumerate}[(1)]
\item
$0\leq U_f(\rho)\leq 1-\frac{1}{d}$. $U_f(\rho)=0$ if and only if $\rho$ is certain, $U_f(\rho)=1-\frac{1}{d}$ if and only if $\rho$ is maximally uncertain.
\item
$U_f(\rho)$ is a concave function of $\rho$, $U_f(p \rho_1 +(1-p) \rho_2)\geq  p U_f(\rho_1) +(1-p) U_f(\rho_2)$ for $0\leq p \leq 1$.
\item
$U_f(\rho)$ is invariant under the action of unitary operation $W$, $U_f(W\rho W^\dagger)=U_f(\rho)$.
\item
$U_f(\rho)$ is decreasing under partial trace in the sense that $U_f(\rho_{A(B)})\leq U_f(\rho_{AB})$ for bipartite quantum state $\rho_{AB}$ with reduced states $\rho_{A(B)}={\rm Tr}_{B(A)}\rho_{AB}$.
\item
$U_f(\rho_1)U_f(\rho_2)=U_f(\rho_1)+U_f(\rho_2)-U_f(\rho_1\otimes\rho_2)$ for any quantum states $\rho_1$ and $\rho_2$.
\end{enumerate}

Now we prove these properties item by item. The first item is by $\frac{1}{d}\leq \max_i {\rho_{ii}}\leq 1$ for any quantum state $\rho=\sum_{i,j} \rho_{ij} |i\rangle\langle j|$.
The second item is because the maximum function is a convex function.
The third item is because under the projective measurement $\Pi=\{\Pi_i\}_{i=0}^{d-1}$ with $\Pi_i=|i\rangle\langle i|$, $\Pi(W\rho W^\dagger)=\Pi(\rho)$ for any unitary operation $W$ and quantum state $\rho$.
The fourth item is because $U_f(\rho_{A})=1-\max_i \sum_j \rho_{ij,ij}\leq 1-\max_{i,j} \rho_{ij,ij} = U_f(\rho_{AB})$ for any bipartite quantum state $\rho=\sum_{i,j,k,l} \rho_{ij,kl} |ij \rangle\langle kl|$.
The fifth item is because $U_f(\rho_1)U_f(\rho_2)=(1-\max_{\sigma_1\in\mathcal{C}} F(\rho_1,\sigma_1))(1-\max_{\sigma_2\in\mathcal{C}} F(\rho_2,\sigma_2))=(1-\max_{\sigma_1\in\mathcal{C}} F(\rho_1,\sigma_1))+(1-\max_{\sigma_2\in\mathcal{C}} F(\rho_2,\sigma_2))- (1-\max_{\sigma_1\in\mathcal{C}}\max_{\sigma_2\in\mathcal{C}} U_f(\rho_1,\sigma_1) U_f(\rho_2,\sigma_2))=U_f(\rho_1)+U_f(\rho_2)- U_f(\rho_1\otimes\rho_2)$.

Recall that the geometric coherence defined as
\begin{eqnarray}
C_f(\rho)=1-\max_{\sigma\in\mathcal{I}} F(\rho,\sigma),
\end{eqnarray}
with the maximum running over all incoherent states, is a coherence measure \cite{A. Streltsov2015}. Let
$D_f(\rho)=\max_{\sigma\in\mathcal{I}} F(\rho,\sigma)- \max_{\sigma\in\mathcal{C}} F(\rho,\sigma)$, then the geometric uncertainty can be decomposed as
\begin{eqnarray}
U_f(\rho)=D_f(\rho) + C_f(\rho),
\end{eqnarray}
with $D_f(\rho)$ quantifying the classical uncertainty.
For any pure state $|\psi\rangle$, the classical uncertainty $D_f(|\psi\rangle)$ vanishes and the total uncertainty coincides with quantum uncertainty $U_f(|\psi\rangle)=C_f(|\psi\rangle)$.
For any incoherent state $\rho=\sum_i \rho_{ii} |i\rangle \langle i|$, the quantum uncertainty $C_f(\rho)$ vanishes and the total uncertainty coincides with classical uncertainty $U_f(\rho)=D_f(\rho)$.

\section{Conclusions and discussions}

To conclude, we have studied the uncertainty as the intrinsic property of quantum states systematically with respect to the projective measurement. In this context, some basic concepts are reformulated and some
new results are obtained. We have proved a necessary and sufficient condition for the uncertainty preserving operation and characterized the quantum states with maximal uncertainty. We have also shown a universal decomposition of total uncertainty into quantum uncertainty and classical uncertainty for all uncertainty measures both formally and quantitatively. The unified relation among uncertainty, coherence, coherence of assistance is further established.
Last but not the least, we have proposed a new uncertainty measure called the geometric uncertainty. Overall, this study of the uncertainty as the intrinsic property of quantum states may shed light on the resource theory and its application. On the other hand, it may be also useful for the interpretation of the uncertainty relation.

Here we should emphasize that in this paper we consider the uncertainty of quantum states with respect to the projective measurement. For instance, in qubit systems, under the reference basis $\{|0\rangle, |1\rangle\}$, the pure state $|0\rangle$ is certain and contains null uncertainty. But under the orthonormal basis $\{|+\rangle, |-\rangle\}$ with $|+\rangle=\frac{1}{\sqrt{2}} (|0\rangle+|1\rangle)$ and $|-\rangle=\frac{1}{\sqrt{2}} (|0\rangle-|1\rangle)$, the pure state $|0\rangle=\frac{1}{\sqrt{2}} (|+\rangle+|-\rangle)$ is maximally uncertain and contains the maximal uncertainty in qubit systems. Therefore the results derived here are basis dependent. It also deduces the famous uncertainty relation which refers to the constraints on the uncertainty of the quantum state under different measurement bases.

Furthermore, there are some interesting problems
which can be further considered.
The first question is the relation between uncertainty and other quantumness of quantum states. The relations between the uncertainty and the entanglement \cite{M. Berta},
the partial coherence (quantum uncertainty) and quantum correlation \cite{Xiong} have been revealed previously.
More than that, the uncertainty, coherence and entanglement can be quantified by geometric uncertainty, geometric coherence, and geometric entanglement \cite{T.C.Wei,Streltsov2010} consistently in terms of the fidelity.
And the geometric coherence amounts to
the maximum bipartite geometric entanglement that can be generated via incoherent operations
applied to the system and an incoherent ancilla \cite{A. Streltsov2015}.
So the relation among uncertainty, coherence and entanglement will be more clear if the conversion between geometric uncertainty and geometric coherence is clarified.
Second, instead of the projective measurements, one may consider the uncertainty with respect to the  L\"{u}ders measurement or POVM, which may unify coherence and uncertainty in a more extensive context.
Third, we have proved that the coherence of assistance is no more than the uncertainty. This implies the coherence obtained in the one-way coherence distillation in one-copy setting is no more than the total uncertainty. Does the regularized coherence of assistance $C_a^{\infty}(\rho) =\lim_{n\rightarrow \infty} \frac{1}{n} C_a(\rho^{\otimes n})$ reaches the total uncertainty, that is  $C_a^{\infty}(\rho)=U(\rho)$? The physical explanation is whether the coherence obtained in the one-way coherence distillation in many-copy setting can reach the uncertainty or not.

\section{Acknowledgement}
We thank Shunlong Luo and Nan Li for discussions and valuable suggestions.
This work is supported by the National Natural Science Foundation of
China under grant Nos. 12171044.

\section{Data availability statement}
All data that support the findings of this study are included within the article (and any supplementary files).

\section{Appendices}

\subsection{The proof of Theorem \ref{th certain operation}.}\label{app cer}

\begin{proof}
For any certain quantum operations $\Lambda=\sum_l K_l (\cdot) K_l^\dagger$ with $\sum_l K_l^\dagger K_l=I$, and for any given certain state $|i\rangle\langle i|$, it has $\Lambda(|i\rangle\langle i|) \in {\cal C}$. Without loss of generality, we suppose
$\Lambda(|i\rangle\langle i|)=|g(i)\rangle\langle g(i)|$ with $g(i)\in \{i\}$.
Therefore $K_l (|i\rangle\langle i|) K_l^\dagger=p_{il} |g(i)\rangle\langle g(i)|$
and $K_l |i\rangle=\sqrt{p_{il}} e^{{\rm i}\theta_{il}}|g(i)\rangle$ for any $l$. When $i$ runs all over $\{i\}_{i=0}^{d-1}$, we have $K_l=\sum_i \sqrt{p_{il}} e^{{\rm i}\theta_{il}} |g(i)\rangle \langle i|$.
Since $\sum_l K_l^\dagger K_l=I$, so $\sum_l \sum_{i^\prime} \sum_i \sqrt{p_{i^\prime l} p_{il}}  e^{{\rm i}(\theta_{il} -\theta_{i^\prime l})} \langle g(i^\prime)|  g(i)\rangle |i^\prime\rangle \langle i|=I$. So we have
$\sum_l p_{il}=1$ for all $i$, and $\sum_l  \sqrt{p_{i^\prime l} p_{il}}  e^{{\rm i}(\theta_{il} -\theta_{i^\prime l})}  \langle g(i^\prime)|  g(i)\rangle =0$ for $i\neq i^\prime$.
\end{proof}

\subsection{The nonnegative symmetric concave function}\label{app sym con f}

The nonnegative symmetric concave function defined on the probability simplex $\Omega=\{\vec{x}=(x_0,x_1,\cdots,x_{d-1})^T|\sum_{i=0}^{d-1} x_i =1 \ \text {and} \ x_i\geq0 \}$ is a function such that it is
\begin{itemize}
\item $f((1,0,\cdots,0)^T)=0$.
\item invariant under any permutation transformation $P_{\pi}$, $f(P_{\pi}\vec{x})=f(\vec{x})$.
\item concave, $f(\lambda \vec{x} +(1-\lambda) \vec{y})\geq \lambda f(\vec{x}) +(1-\lambda) f(\vec{y})$ for any $0\leq \lambda \leq 1$ and $\vec{x}, \vec{y}\in \Omega$.
\end{itemize}
To guarantee the uncertainty (coherence) in the uncertain (coherent) states is positive, it is reasonable to demand the
\begin{itemize}
\item $f(\vec{x})=0$ if and only if $\vec{x}=P_{\pi}(1,0,\cdots,0)^T$ for some permutation transformation $P_{\pi}$.
\end{itemize}
Therefore $U(\rho)=0$ ($C(\rho)=0$) if and only if $\rho$ is certain (incoherent). Throughout this paper, when we say the nonnegative symmetric concave function, we default it satisfying this additional condition.

\subsection{The proof of Theorem \ref{th uncertainty pres}.}\label{app uncer pre}
\begin{proof}
On one hand, suppose the quantum operation $\Lambda$ preserving uncertainty for every quantum state $\rho$, $U(\Lambda(\rho))=U(\rho)$. First we assume $\rho$ is certain, then $U(\rho)=0$. Therefore $U(\Lambda(\rho))=0$, which implies $\Lambda(\rho)$ is a certain state. This requires that $\Lambda(\rho)$ is certain for all certain states $\rho$, so $\Lambda$ is a certain operation.
By Theorem \ref{th certain operation}, we suppose  $\Lambda=\sum_l K_l (\cdot) K_l^\dagger$ with $K_l=\sum_i \sqrt{p_{il}} e^{{\rm i}\theta_{il}} |g(i)\rangle \langle i|$. For quantum state $\rho=\sum_{i,j=0}^1 \rho_{ij} |i\rangle\langle j|$,
its uncertainty is $U(\rho)=f(\rho_{00},\rho_{11},0,\cdots,0)$. After the action of quantum operation $\Lambda$, the resulted state is $\Lambda(\rho) = \sum_{i,j=0}^1 \sum_l \sqrt{p_{il}p_{jl}} e^{{\rm i}(\theta_{il}-\theta_{jl})} \rho_{ij} |g(i)\rangle \langle g(j)|$. Due to the arbitrariness of $\rho$, we have $g(i)\neq g(j)$ for any $i\neq j$. Therefore $g$ is a permutation in $\{i\}$. This implies the Kraus operators $K_l$ is in form of $K_l=D_l P_{\pi}$ with some diagonal matrices $D_l$ and permutation $P_{\pi}$.

On the other hand, for the quantum operation $\Lambda=\sum_l K_l(\cdot)K_l^\dagger$ with Kraus operators $K_l= D_lP_{\pi}$ with diagonal matrices $D_l$ such that $\sum_l D_l^\dagger D_l=I$ and some permutation $P_{\pi}$, it is easy to verify that $\Pi(\Lambda(\rho))=\sum_{i} \rho_{ii} |\pi(i)\rangle \langle \pi(i)|$, therefore
$U(\Lambda(\rho))=U(\rho)$.
\end{proof}

\subsection{The proof of Theorem \ref{th maxi uncer}}\label{app maxi uncer}

\begin{proof}
For any quantum state $\rho=\sum_{ij} \rho_{ij} |i\rangle \langle j|$,
it satisfies $(1/d, 1/d, \cdots, 1/d) \prec (\rho_{00}, \rho_{11},\cdots, \rho_{d-1,d-1})$. Hence $f(\rho_{00}, \rho_{11},\cdots, \rho_{d-1,d-1})\leq f(1/d, 1/d, \cdots, 1/d)$ for any Schur concave function $f$ \cite{Bhatia}, which implies $U(\rho)\leq f(1/d, 1/d, \cdots, 1/d)$.
Therefore, $\rho$ is maximally uncertain if and only if $\rho_{ii}=1/d$ for $i=0,1,\cdots, d-1$.
\end{proof}


\subsection{Proof of Theorem \ref{th relation}.}\label{app rel}
\begin{proof}
Here we only need to prove the right hand side inequality $C_a(\rho) \leq U(\rho)$.
In fact, for any nonnegative  symmetric concave function $f$ and any pure state $|\psi\rangle$, the uncertainty coincides with quantum coherence as $U(|\psi\rangle)=C(|\psi\rangle)=f(|\psi_0|^2,|\psi_1|^2,\cdots,|\psi_{d-1}|^2 )$. For any mixed state $\rho=\sum_{i,j} \rho_{ij} |i\rangle\langle j|$, suppose $\rho=\sum_k p_k |\psi_k
\rangle \langle \psi_k|$ with $|\psi_k\rangle=\sum_{i} \psi_i^{(k)} |i\rangle$ is any pure-state decomposition of $\rho$. Then we have
\begin{equation}\label{cohandun}
\begin{array}{rcl}
C(\rho)&\leq& \sum_k p_k C(|\psi_k\rangle) \\&=& \sum_k p_k f(|\psi_0^{(k)}|^2, |\psi_1^{(k)}|^2,\cdots, |\psi_{d-1}^{(k)}|^2) \\
&\leq&  f(\sum_k p_k|\psi_0^{(k)}|^2, \sum_k p_k|\psi_1^{(k)}|^2,\cdots, \sum_k p_k|\psi_{d-1}^{(k)}|^2)\\
&=&f(\rho_{00}, \rho_{11},\cdots, \rho_{d-1,d-1})\\
&=&U(\rho),
\end{array}
\end{equation}
where the first inequality is the convexity of coherence measures and the second inequality is the concavity of function $f$. So we get $C_a(\rho) \leq U(\rho)$.
\end{proof}



\subsection{The proof of Theorem \ref{th uf}.}\label{app uncer fie}
\begin{proof}
For any quantum state $\rho=\sum_{i,j} \rho_{ij} |i\rangle \langle j|$, it has
$F(\rho,|i\rangle\langle i|)=[Tr(\sqrt{\rho} |i\rangle\langle i|\sqrt{\rho})^\frac{1}{2}]^2=[\frac{1}{\sqrt{\rho_{ii}}} Tr \sqrt{\rho} |i\rangle\langle i|\sqrt{\rho}]^2=[\sqrt{\rho_{ii}}]^2=\rho_{ii}$, therefore
\begin{eqnarray}\label{eq ana uf}
U_f(\rho)=1-\max_{i} F(\rho,|i\rangle\langle i|)=1-\max_{i} \rho_{ii}.
\end{eqnarray}
Since $f(\rho_{00}, \rho_{11},\cdots, \rho_{d-1,d-1})=1-\max_{i} \rho_{ii}$ is a nonnegative symmetric concave function, it demonstrates that $U_f$ is a bona fide uncertainty measure.
\end{proof}

\end{document}